\documentclass[aps,prl,preprint,groupedaddress,showpacs,showkeys]{revtex4}

\usepackage{graphics}
\usepackage{graphicx}

\begin{document}

\title{The Origin of Doping in Quasi-Free Standing Graphene on Silicon Carbide}

\author{J.~Ristein}
\affiliation{Lehrstuhl f\"ur Technische Physik, Universit\"at Erlangen-N\"urnberg, Erwin-Rommel-Str.1, 91058 Erlangen, Germany}

\author{S.~Mammadov}
\affiliation{Lehrstuhl f\"ur Technische Physik, Universit\"at Erlangen-N\"urnberg, Erwin-Rommel-Str.1, 91058 Erlangen, Germany}

\author{Th.~Seyller}
\email[Corresponding author. Email: ]{thomas.seyller@physik.uni-erlangen.de}
\affiliation{Lehrstuhl f\"ur Technische Physik, Universit\"at Erlangen-N\"urnberg, Erwin-Rommel-Str.1, 91058 Erlangen, Germany}

\date{\today}

\begin{abstract}
We explain the robust p-type doping observed for quasi-free standing graphene on hexagonal silicon carbide by the spontaneous polarization of the substrate. This mechanism is based on a bulk property of SiC, unavoidable for any hexagonal polytype of the material and independent of any details of the interface formation. We show that sign and magnitude of the polarization are in perfect agreement with the doping level observed in the graphene layer. With this mechanism, models based on hypothetical acceptor-type defects as they are discussed so far are obsolete. The n-type doping of epitaxial graphene is explained conventionally by donor-like states associated with the buffer layer and its interface to the substrate which overcompensate the polarization doping.
\end{abstract}



\maketitle

The basis for the unique electronic and optical properties of graphene is the linear dispersion relation of the $\pi$-electrons which is responsible for Dirac type quasi-particles with many unusual properties. The band structure in the relevant energy range is made up by double cones in the corners of the two-dimensional hexagonal Brilloun zone; their opening angle is determined by the slope $v_{F}=\frac{d\omega}{dk}$  of the dispersion relation called the Fermi velocity which is an intrinsic material parameter. The origin of these so-called Dirac cones defines the Fermi energy in an isolated and intrinsic graphene layer. At finite temperatures, the reservoir of mobile charge carriers is due to thermal excitation of equal concentrations $n_{0}$ and $p_{0}$ of electrons and holes. Evaluation of the Fermi statistics yields a value of $n_{0}=\frac{\pi k_{B}^{2}}{6\hbar v_{F}^{2}}T^{2} $ for this intrinsic charge carrier concentration at temperature $T$ where $k_{B}$ and  $\hbar$  are the Boltzmann constant and Planck's constant, respectively\cite{Ristein2010}. For room temperature, this amounts to $n_{0}=6.1\cdot 10^{10}$~$\mathrm{cm^{-2}}$ when $v_{F}=1.1\cdot 10^{6}$~$\mathrm{ms^{-1}}$ is taken for the Fermi velocity \cite{Bostwick2007,Ohta2007a}.

Epitaxial graphene on hexagonal silicon carbide is never found to follow this expectation \cite{First2010a}. Graphene on SiC(0001) usually reveals a pronounced n-type conductivity with a Fermi level position about 0.45 eV above the Dirac energy and corresponding electron concentrations of $1.2\cdot 10^{13}$~$\mathrm{cm^{-2}}$, i.e. a value more than two orders of magnitude larger than $n_{0}$. For 6H-SiC (0001) the band alignment between this substrate material and the graphene layer is asymmetric with the Dirac energy 0.75 eV below the SiC conduction band minimum (CBM) and 2.25 eV above the valence band maximum (VBM).

The n-type character of the graphene can be explained by donor-like states at the SiC/graphene interface. This interface is sketched in Fig. 1(a) along with a band diagram which will be discussed below. The graphene layer is sitting on top of a buffer layer that intimately resembles the surface layer of the so-called  $6\sqrt{3}\times 6\sqrt{3} R30^{\circ}$ reconstruction of SiC(0001) that is almost commensurate with the graphene lattice \cite{Emtsev2008}. The buffer layer is partially bound to the silicon atoms of the substrate and is considerably distorted as compared to graphene. Photoelectron spectroscopy shows surface states that can be associated with carbon or silicon dangling bonds superimposed on a broad density of states for the buffer layer that extends up to the Fermi level\cite{Emtsev2008}. The complete ensemble of electronic states associated with the buffer layer and the SiC surface can be combined in an interface density of states with a charge neutrality level $E_{0}$ as shown schematically in the band diagram of Fig. 1(a). The n-type character of epitaxial graphene on SiC(0001) is most plausibly explained by electron transfer from these interface density of states to the graphene layer.

\begin{figure}
\centering
\includegraphics[width=\textwidth]{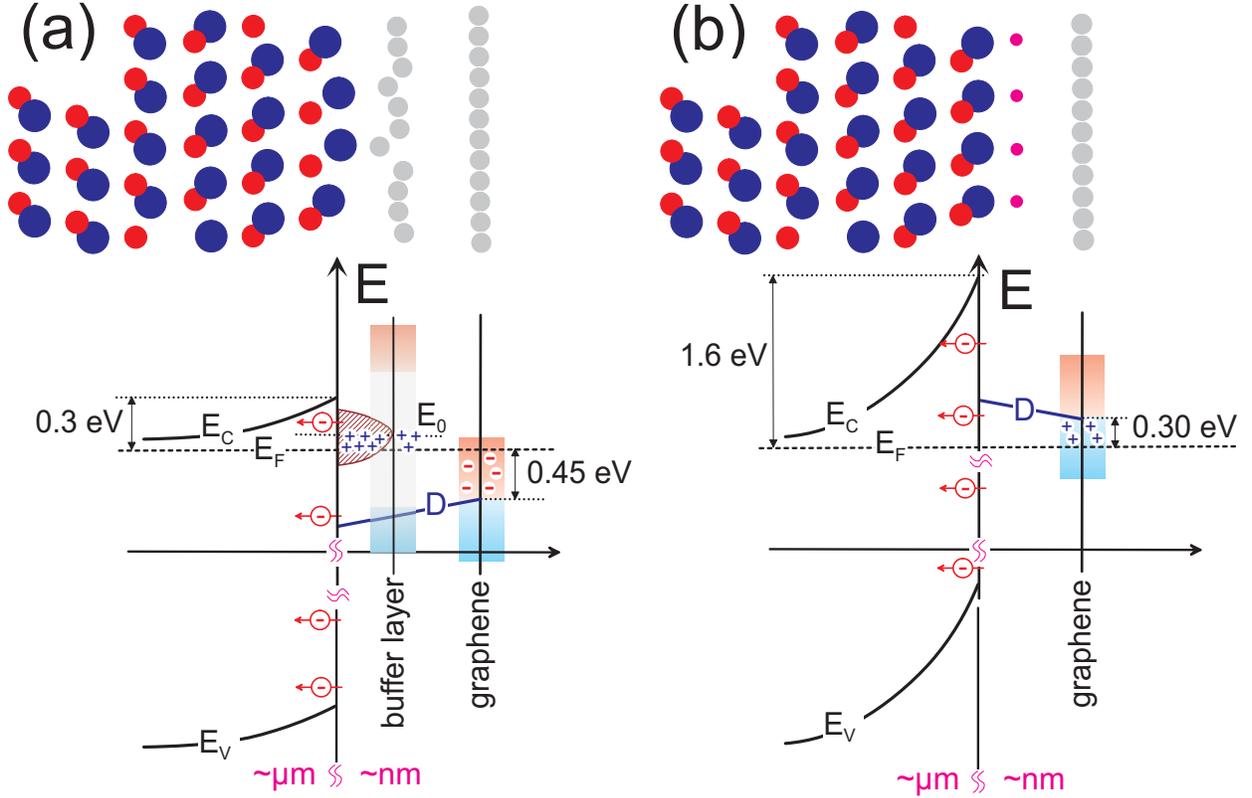}
\caption{Sketch and schematic band diagram for the epitaxial (a) and the quasi-free standing (b) graphene interface with 6H-SiC(0001). Large and small circles represent silicon and graphene atoms, respectively, and the very small circles in (b) stand for hydrogen. The polarization vector of the SiC is indicated by the little arrows at the interface along with the negative polarization charge discussed in the text. D denotes the Dirac energy. For the band bending information from photoelectron spectroscopy \cite{Speck2010} has been taken into account.}
\end{figure}

We should mention at this point that Kopylov et al.\cite{Falko2010} have recently discussed this type of doping for monolayer and bilayer graphene on a phenomenological base along with an alternative mechanism based on classical donors in the SiC substrate. However, in order to explain electron densities as measured in epitaxial graphene, a bulk doping level of more than  $10^{19}$~$\mathrm{cm^{-3}}$  is needed which is incompatible with the type of SiC substrates commonly used. Moreover, the doping mechanism changes and is even inverted when the interface between SiC and graphene is passivated by hydrogen (see below). Both arguments suggest that interface rather than bulk states act as donors. We note in passing a very asymmetric alignment of the charge neutrality level $E_{0}$ of the interface density of states which is around 0.3 eV below the CBM and 2.7 eV above the VBM of 6H-SiC. 

The situation becomes even more mysterious when a quasi-free standing monolayer graphene (QFMLG) on hexagonal SiC(0001) is prepared. This hetero system forms when the carbon rich $(6\sqrt{3}\times 6\sqrt{3})R30^{\circ}$ reconstructed top layer of the SiC(0001) surface is delaminated by hydrogen intercalation \cite{Riedl2009}. With this technique, the carbon-silicon bonds to the substrate are released and the resulting silicon dangling bonds are saturated by hydrogen with remarkable perfection \cite{Speck2011}. The Si-H bonds are electronically inactive, and in principle a defect-free interface between SiC and graphene is expected. Consequently, charge transfer to the graphene layer is now restricted to the depletion of the bulk doping of SiC which is insufficient for any substantial doping as has been argued above. Hence, for the n-type and semi-insulating substrate material usually adopted for quasi-free standing graphene on SiC(0001) very little n-type doping of the graphene layer by the substrate and a Fermi level in close vicinity of the Dirac energy is expected. The experimental observation, however, is the exact opposite: instead of mild n-type doping QFMLG on 6H-SiC(0001) exhibits a strong p-type conductivity with the Fermi level 0.30 eV below the Dirac energy and hole densities of about $5,5\cdot 10^{12}$~$\mathrm{cm^{-2}}$, i.e. comparable to the electron densities observed for the epitaxial graphene. A mechanism along the classical lines would require a reservoir of acceptor-like defects now of a comparable capacity, although the interface is passivated by the hydrogen termination of the substrate. All classical models, e.g. involving the anti-bonding orbitals of Si-H bonds or acceptor-like hypothetical defects of this interface as doping reservoir are implausible. The question, where the robust p-type doping of hydrogen intercalated graphene on hexagonal SiC comes from, is therefore a mystery so far and the unravelling of this mystery is the main subject of this paper.

We suggest that the intrinsic spontaneous polarization of hexagonal SiC is responsible for the p-type doping of the quasi-free standing graphene layer on top of it. To elucidate this mechanism somewhat, let us briefly reconsider (i) Coulombs fundamental law of electrodynamics $\vec{\nabla}\cdot\vec{D}=\rho$ linking the dielectric displacement $\vec{D}$  to the space charge density $\rho$ and (ii) the macroscopic linear relation between electric field $\vec{E}$  and dielectric displacement in matter:
\begin{equation}
\vec{D}=\varepsilon_{0}\vec{E}+\vec{P}=\varepsilon_{0}\vec{E}+\vec{P_{0}}+\chi\varepsilon_{0}\vec{E}=\varepsilon_{0}(1+\chi)\vec{E}+\vec{P_{0}}
\end{equation}
$\varepsilon_{0}$ is the vacuum permeability, $\chi=1-\varepsilon$ the electric susceptibility, and $\vec{P}$ the dipole density or polarization in the material under consideration which may have a spontaneous $\vec{P_{0}}$ and a field induced component $\chi\varepsilon_{0}\vec{E}$. If we apply both equations to a layer stack system of two materials with dielectric constants $\varepsilon_{1}$ and $\varepsilon_{2}$ the change $\Delta(\varepsilon E)=\varepsilon_{2}E_{2}-\varepsilon_{1}E_{1}$ of the weighted (normal) electric field components across the interface can be calculated yielding:
\begin{equation}
\Delta(\varepsilon E)=\sigma -\Delta P_{0}
\end{equation}

For non-ferroelectric materials $P_{0}$ is zero and (2) reduces to the familiar relation between the field change and the integrated space charge density $\sigma$ (in units of charge per area) which holds for any arbitrarily selected range of the layer stack system. When ferroelectrics like the hexagonal polytypes of SiC are involved, however, (2) shows that a difference in spontaneous polarization across an interface has the same effect as a negative sheet charge at the interface. Thus,  $-\Delta P_{0}$ can be associated with a pseudo charge which we may call polarization charge in the following.

Although becoming important at interfaces only, the spontaneous polarization is a bulk property of the materials involved, and its change across a specific interface is independent of any structural or geometric details of that interface. Specifically, when a slab of ferroelectric material is embedded between non-ferroelectric layers -also vacuum may serve as the latter-, the two surfaces inevitably carry a polarization charge with identical magnitude and opposite sign on the two interfaces that is an intrinsic bulk property of the ferroelectric material. For any electronic material brought into contact with the ferroelectric, this polarization charge is fully equivalent to a sheet of charged dopants on the ferroelectric side of the interface, however without the necessity of any real quantum mechanical states associated with those charges. We will show in the following that sign and magnitude of the spontaneous polarization of 6H-SiC(0001) is perfectly consistent with the acceptor concentration needed to explain the p-type conductivity of H-intercalated graphene on that surface.

As already indicated above, spontaneous polarization is an intrinsic feature of all polytpes of SiC except for 3C-SiC. For the classical semiconductors Si, Ge, GaAs or diamond this feature is missing due to the cubic symmetry of the material, and therefore it is not considered at hetero interfaces of these semiconductors. It plays an important role, however, in the device physics of GaN and its derivatives which are of wurzite, i.e. hexagonal crystal structure \cite{Ambacher1999}. Amazingly, it is very difficult to address it by experiment. It essentially leads to effective surface charges on the polar surfaces of the semiconductor compounds which are usually compensated completely or in part by internal charge formation due to depletion or accumulation of charge carriers or by external build-up of ionic charge on the surfaces. Only the \emph{modulation} of the effective polarization charge at a surface upon an external perturbation like strain, temperature variation or the presence of optical phonons can be measured \cite{Austerman1963}\cite{Loh1968}\cite{madelung-Springer1979}, but not its base value. As a consequence, no experimental data are available for the polarization of any hexagonal semiconductor crystal. In order to discuss the polarization of SiC polytypes we are therefore restricted to theoretical results.

However, also for theory spontaneous polarization is conceptually difficult to address. The reason is that for any model with periodic boundary conditions, the dipole density $\vec{P}=\frac{1}{V}\int \vec{r}\cdot \rho(\vec{r})dV$ of a system is ill defined since it depends on the arbitrary choice of the unit cell boundaries over which it is evaluated. Although the charge density $\rho(\vec{r})$ alone can be quite reliably and accurately determined by modern density functional techniques, the direct evaluation of the polarization is impossible. A break through on this route was achieved by Posternak at al. in 1990 for BeO \cite{Posternak1990}\cite{Resta1990}. They constructed a periodic arrangement with a supercell containing a certain number of double layers of the wurzite form together with a number of double layers of the zinc blende form. The latter one has zero polarization by symmetry. When calculating the potential along this periodic superstructure by density functional theory, the Born-von Karman boundary conditions automatically eliminate the average electric field. However, the discontinuity of this field at the boundaries between the wurzite and the zinc blende slab is independent of this artefact. This field discontinuity can directly be translated by (2) to the change in polarization. Because $P_{0}$ is known to be zero in the zinc blende slabs, $P_{0}$ can be evaluated by this trick for the wurzite parts of the heterostructure. This quite ingenious approach was later taken by Qteish, Heine and Needs \cite{Needs1992}\cite{Needs1993} to calculate the spontaneous polarization of various SiC polytypes by density functional theory. They presented results for 2H (wurzite), 6H-, and 8H-SiC.

Part of the spontaneous polarization can be ascribed to the relative elongation of the longitudinal Si-C bonds along the $\vec{c}$-axis and the contraction of the bonds transvers to it. For the wurzite crystal structure (2H-SiC) with only two double layers in the unit cell, this distortion is characterized by the longitudinal bond length in units of the hexagonal lattice parameter $\vec{c}$, denoted as internal parameter $u$. When bond lengths and angles are chosen the same as in the cubic polytype (zinc blende) and only the stacking sequence is changed form staggered to eclipsed the so-called ideal wurzite structure with $u=3/8$ is achieved. The projection of the transverse bonds on the   $\vec{c}$-axis is then $1/8c$, and since they are three times as many as the longitudinal bonds and point backwards relative to them, a simple geometric evaluation of the dipole density based on equally polarized bonds results in zero. In real cases, the internal parameter $u$ for wurzite alloys is usually between 0.2 and 2$\%$ larger than 3/8 \cite{Bernardini1997}. By taking only this purely geometric effect into account, a negative polarization (polarization vector opposing the conventional $\vec{c}$-axis direction) for hexagonal SiC corresponding to negative effective polarization charge on the Si-terminated (0001) surface is expected. However, even when this geometric relaxation is suppressed in DFT calculations by choosing all Si-C bond lengths identical in the hexagonal as well as in the cubic slabs of the model system (i.e. modelling the ideal wurzite structure or its analog), the loss of the cubic symmetry alone induces a relaxation of the space charge density with a contribution to the polarization that has the same sign but an even larger magnitude than the ionic relaxation contribution for 2H-SiC. ($-3.33\cdot 10^{-2}$~$\mathrm{C/m^{2}}$ compared to $-0.99\cdot 10^{-2}$~$\mathrm{C/m^{2}}$) \cite{Needs1993}\footnote[1]{Following the commonly used sign convention for the $\vec{c}$ axis direction in hexagonal ferroelectrics, the spontaneous polarization $\vec{P_{0}}$ of SiC polytypes as calculated by Qteish et. al is opposite to $\vec{c}$. Nevertheless the authors quantify their results in Refs. \cite{Needs1993,Needs1992} with positive $P_{0}$ values.  Keep this in mind when comparing their results with other literature!}. This is quite in contrast to wurzite BeO that has first been studied as a ferroelectric model substance \cite{Posternak1990}. Here, the geometric contribution to the polarization is the dominating one.

Unfortunately, and probably due to the computational demand of an ab initio and self consistent quantum mechanical modelling, only the 2H polytype of SiC with the smallest unit cell has been treated by the approach outlined above \cite{Needs1993}. For 6H- and 8H-SiC the spontaneous polarization was only calculated for an idealized hexagonal crystal structure with bond lengths and angles like in cubic SiC. The contribution of the ionic relaxation of the real polytypes relative to this structure is thus not included in the calculations. We may take the ratio $c/a$ of the longitudinal and transverse hexagonal lattice parameters as relative measure for this ionic relaxation. It deviates from the corresponding values for the idealized structures by +0.49$\%$ for 2H-SiC, +0.15$\%$ for 4H-SiC, and +0.16$\%$ for 6H-SiC \cite{Park1994}. The effect of ionic relaxation is thus about three times larger in 2H-SiC than in the other polytypes, and since it contributes only 23$\%$ to the spontaneous polarization even in the former case we may neglect it within the uncertainties of the calculations for 4H and 6H polytypes. For 6H-SiC adopted usually as a substrate for quasi-free standing graphene, the spontaneous polarization then is $P_{0}=-9.49\cdot 10^{-3}$~$\mathrm{C/m^{2}}$. As outlined above, this value is equivalent to a sheet of negatively charged donors on the Si face of the substrate with density $\left| P_{0} \right|/e=5.9\cdot 10^{12}$~$\mathrm{cm^{-2}}$. This deviates by less than 10$\%$ from the hole density in the quasi-free standing graphene layer measured on top of it (see above).

It should be mentioned at this point that Davydov and Troshin \cite{Davydov2007} have reviewed in 2007 the calculations of spontaneous polarization in SiC polytypes performed by different methods. They find a considerable scatter of the results that ranges for 2H-SiC from $-1.11$ to $-4.32\cdot 10^{-2}$~$\mathrm{C/m^{2}}$. This indicates a substantial uncertainty also for the theoretical value of 6H-SiC cited above. The agreement between the graphene hole density and the calculated polarization may therefore be somewhat fortuitous.

We conclude that spontaneous polarization of the hexagonal SiC substrate is directly responsible for the doping of quasi-free standing graphene on its (0001) surface. The spontaneous polarization creates a pseudo-acceptor layer which is fully equivalent to real acceptors. Being a bulk property of the substrate, it constitutes an invariant pseudo-sheet charge density only characteristic for the respective hexagonal polytype of the substrate material. We have illustrated this doping mechanism of quasi-free standing graphene on hexagonal SiC(0001) in the band diagram of Fig. 1(b).

The pseudo-polarization charge will be present for epitaxial graphene layers as well, but in this case it is obviously overcompensated by doping due to the interface density of states as explained above. Thus, when the spontaneous polarization of the substrate is properly taken into account, this density of states must account for both $-P_{0}/e$ and the observed electron density $n$ in the epitaxial graphene layer.

There are a number of obvious tests of the polarization doping model suggested here: (i) preparing quasi-free standing graphene with a defect-free interface on the carbon-terminated faces of the hexagonal SiC substrates should lead to n-type doping; (ii) on cubic SiC it should lead to undoped graphene layers; (iii) varying the hexagonality of the SiC polytype should increase the doping level proportionally since the spontaneous polarization is mainly induced at the inversion of the stacking sequence of the double layers \cite{Needs1993}. For 4H-SiC, for example, the polarization is expected to be 6/4 times larger than for 6H-SiC. For most of these tests, the preparation of the corresponding defect free interfaces has not been realized to date. Only for the silicon terminated surface of 3C-SiC Coletti et al. \cite{Coletti-3C-2011} have recently reported on quasi-free standing graphene. In fact, the p-type doping as observed for the 6H-SiC is absent, and a mild n-type doping with an electron density of about $1.3\cdot10^{12}$~$\mathrm{cm^{-2}}$ is found instead that can easily be explained by residual defects at the interface. We interpret the vanishing of the p-type character for QFMLG when going from the hexagonal 6H to the cubic 3C polytype as a very strong argument in support of the polarization doping model suggested in this paper.

The authors are grateful to Prof. Lothar Ley for many fruitful and stimulating discussions. The work was supported by the Deutsche Forschungsgemeinschaft and by the ESF through the EUROCORES program EuroGRAPHENE.


\begin{thebibliography}{5}

\bibitem{Ristein2010} J. Ristein et al., J. Phys. D: Appl. Phys. \textbf{43}, 345303 (2010).

\bibitem{Bostwick2007} A. Bostwick et al., Nat. Phys. \textbf{3} 36\textendash40 (2007).

\bibitem{Ohta2007a} T. Ohta et al., Phys. Rev. Lett. \textbf{98},  206802 (2007).

\bibitem{First2010a} P.~N. First et al., MRS Bull. \textbf{35}, 296 (2010)

\bibitem{Emtsev2008} K.V. Emtsev, F. Speck, Th. Seyller, J.D. Riley and L Ley, Phys. Rev B \textbf{77}, 155303 (2008).

\bibitem{Speck2010} F. Speck et al., Mater. Sci. Forum \textbf{645-648}, 629 (2010).

\bibitem{Falko2010} S. Kopylov, A. Tzalenchuk, S. Kubatkin, and V.I. Falko, Appl. Phys. Lett. \textbf{97}, 112109 (2010).

\bibitem{Riedl2009} C. Riedl, C.Coletti et al., Phys. Rev. Lett. \textbf{103}, 246804 (2009).

\bibitem{Speck2011} F. Speck, J. Jobst et. al., Appl.Phys. Lett. \textbf{99}, 122106 (2011).

\bibitem{Ambacher1999}O. Ambacher et. al., J. Appl. Phys. \textbf{85}, 3222 (1999).

\bibitem{Austerman1963} S.B. Austerman, D.A. Berlincour, and H.H.A. Krueger, J. Appl. Phys. \textbf{34}, 339 (1963).

\bibitem{Loh1968}E. Loh, Phys. Rev. \textbf{166}, 673 (1968).

\bibitem{madelung-Springer1979}\textit{Elastic, Piezoelectric and Related Constants of Crystals}, O. Madelung (ed.), Landolt-B\"{o}rnstein Series, Vol III (Springer, Berlin, 1979).

\bibitem{Posternak1990} M. Posternak et al., Phys. Rev. Lett. \textbf{64}, 1777 (1990).

\bibitem{Resta1990} R. Resta et al., Ferroelectrics \textbf{111}, 15 (1990).

\bibitem{Needs1992} Q. Qteish, V. Heine, and R.J. Needs, Phys. Rev B \textbf{45}, 6376 (1992).

\bibitem{Needs1993} Q. Qteish, V. Heine, and R.J. Needs, Physica B \textbf{185}, 366 (1993).

\bibitem{Bernardini1997} F. Bernardini, V. Fiorentini, D. Vanderbilt, Phys. Rev. B \textbf{56}, R10024 (1997).

\bibitem{Park1994} C.J. Park et al., Phys. Rev. B \textbf{49}, 4485 (1994).

\bibitem{Davydov2007} S. Yu. Davydov and A. V. Troshin, Physics of the Solid State \textbf{40}, 759 (2007).

\bibitem{Coletti-3C-2011} C. Coletti et al., Appl. Phys. Lett. \textbf{99}, 081904 (2011).

\end{thebibliography}
\end{document}